\documentclass[aip,apl,
 amsmath,amssymb,
 reprint,
]{revtex4-1}

\usepackage{graphicx}
\usepackage{dcolumn}
\usepackage{bm}

\usepackage{hyperref}
\hypersetup{hypertex=true,
colorlinks=true,
linkcolor=blue,
anchorcolor=blue,
citecolor=blue}

\usepackage[utf8]{inputenc}
\usepackage[T1]{fontenc}
\usepackage{mathptmx}
\usepackage{etoolbox}
\usepackage{color}
\makeatletter
\def\@email#1#2{%
 \endgroup
 \patchcmd{\titleblock@produce}
  {\frontmatter@RRAPformat}
  {\frontmatter@RRAPformat{\produce@RRAP{*#1\href{mailto:#2}{#2}}}\frontmatter@RRAPformat}
  {}{}
}%
\makeatother
\begin{document}

\preprint{AIP/123-QED}

\title{Electric fields-tuning plasmon and coupled plasmon-phonon modes in monolayer transition metal dichalcogenides}
\author{Chengxiang Zhao}
\affiliation{College of Physics and Information Engineering, Shanxi Normal University, Taiyuan 030031, P. R. China}

\author{Wenjun Zhang}
\affiliation{College of Physics and Information Engineering, Shanxi Normal University, Taiyuan 030031, P. R. China}

\author{Haotong Wang}
\affiliation{College of Physics and Information Engineering, Shanxi Normal University, Taiyuan 030031, P. R. China}

\author{Fangwei Han}
\affiliation{School of Medical Information Engineering, Jining Medical University, Jining 272067, P. R. China}

\author{Haiming Dong}
\thanks{Authors to whom correspondence should be addressed: hmdong@cumt.edu.cn}
\affiliation{School of Materials and Physics, China University of Mining and Technology, Xuzhou 221116, P. R. China}

\date{\today}
\begin{abstract}
We theoretically investigate the electric field-tuning plasmons and plasmon-phonon couplings of two-dimensional (2D) transition metal dichalcogenides (TMDs), such as monolayer MoS$_2$, under the consideration of spin-orbit coupling. It is revealed that the frequencies of plasmons and coupled plasmon-phonon modes originating from electron-electron and electron-phonon interactions can be effectively changed by using applied driving electric fields. Notably, these frequencies exhibit a decreasing trend with an increasing electric field. Moreover, the weak angular dependence of these modes suggests that the driving electric field does not induce significant anisotropy in the plasmon modes. The outcomes of this work demonstrate that the plasmon and coupled plasmon-phonon modes can be tuned not only by manipulating the electron density via the application of a gate voltage but also by tuning the applied driving electric field. These findings hold relevance for facilitating the application of 2D TMDs in optoelectronic devices.
\end{abstract}

\maketitle
\section{Introduction}
In recent years, two-dimensional (2D) materials have garnered substantial attention due to their immense potential in optoelectronic nanodevices. A pivotal phenomenon within these devices is the collective excitation of free electrons, namely plasmons. Plasmons, which are oscillations of electron density within materials, predominantly occur in metals and doped semiconductors. They possess the ability to interact with phonons, photons, and other quasiparticles, thereby enhancing local electromagnetic fields \cite{r1} and manipulating the light \cite{r2}, which has been developed for applications such as biosensors \cite{r3}, superlens imaging \cite{r4}, plasmonic nanolithography \cite{r5}, and so forth recently.

Moreover, phonons-quantized vibrational modes of a material's atomic lattice play a crucial role in determining the optoelectronic performance. They exert an influence on carrier transport, light absorption, and thermal management. The interaction between plasmons and phonons gives rise to intricate phenomena, thereby broadening the scope of applications of 2D materials in optoelectronics. In particular, the plasmon and plasmon-phonon coupling in 2D TMDs, such as monolayer MoS$_2$ (M-MoS$_2$) with a direct band gap and intrinsic spin-orbit coupling (SOC) \cite{r6}, have been investigated theoretically \cite{r7,r8,r9}. Experimental and theoretical studies have confirmed that due to the symmetry breaking at the K and K' points in the Brillouin zone, in M-MoS$_2$, in addition to the low-energy plasmon mode originating from intra-valley transitions, which is proportional to $\sqrt{q}$, there also exists a plasmon mode with higher energy and linear dispersion originating from inter-valley transitions of electrons \cite{re, kk}. The SOC can lead to different
Fermi surfaces at the K and K' points, creating an energy gap in the inter-valley scattering process, and the inter-valley plasmon modes are suppressed \cite{re}. Circularly polarized light can selectively excite the electron transitions in the K and K' valleys respectively \cite{tc, xc,hz,kf}. This selection rule provides a basis for realizing the coupling between plasmons and the valley degree of freedom, thus enabling the realization of optically controlled spin polarization and valley plasmonics devices. Moreover, by combining the spin-orbit coupling with the valley characteristics, multi-component plasmon modes with different properties can be excited by circularly polarized light \cite{ym}.

Based on the special energy band structure, strong spin-orbit coupling, and valley characteristics of monolayer TMDS, they are also excellent polariton materials. For example, due to the excitation of exciton-polaritons in M-MoS$_2$, the spontaneous emission rate of quantum emitters is enhanced by several orders of magnitude compared to the free space value \cite{vdk}. Moreover, there are novel in-plane propagating exciton polaritons in TMDS. Their wavelength localization is two orders of magnitude higher than that of surface plasmon polaritons, and the coupling characteristics of their valley degrees of freedom provide the possibility of realizing directional polariton waves \cite{ie1,ie2}. The hyperbolic exciton-polaritons (HEPs) based on TMDs, coupled with the valley degree of freedom, exhibit the hyperbolic spin-valley Hall effect. These highly confined and valley-polarized HEPs provide a new way to control strong light-matter interactions at the atomic scale \cite{ie3, aj,ie4}. 

Recent studies on plasmon and plasmon-phonon coupling in 2D TMDs have demonstrated that these properties can be modulated by varying the electron densities. Current research efforts predominantly focus on tuning plasmon by controlling the electron density through the application of gate voltage \cite{r10, r11} or chemical doping \cite{r12}. Nevertheless, the large-area gate metal electrode can induce a shielding effect, which has a detrimental impact on the material's electrical transport properties and its interaction with light \cite{r131}.

In optoelectronic devices, a longitudinal driving electric field is of paramount importance for device operation. An essential question thus arises: How do electric fields govern plasmons and their couplings with phonons in a 2D system? This is a critical issue for the efficient operation of such devices. From the perspective of condensed matter physics, electrons can acquire momentum and energy from the electric field, leading to a change in the electron distribution and, consequently, altering the electron behavior under a driving electric field \cite{r13}, that is, the frequency of plasmons resulting from the collective excitation of electrons. Furthermore, the driving electrodes fabricated at the sides of the material have minimal impact on its critical region. Tuning plasmons via a driving electric field can enhance the applicability of 2D electronic systems in optoelectronic devices \cite{r132}. In this study, we explore the impact of the driving electric field on the behavior of plasmons and the interaction between plasmons and phonons in TMDs, considering the spin-orbit coupling. Our findings indicate that both plasmon and coupled plasmon-phonon modes can be effectively tuned by applying a driving electric field. These results are expected to have a profound impact on the performance of 2D electronics.

\section{Theoretical Approaches}
In the present study, we consider a typical 2D device with monolayer TMD (e.g. n-doped MoS$_2$ monolayer sample) on top of a dielectric wafer in the $xy$ sample plane. A source-to-drain electric field $F_x$ is applied along the $x$ direction of the sample plane. The conducting carriers are electrons and the intrinsic energy of electrons near the K-point in MoS$_2$ are \cite{r9}
\begin{equation}
E_{\lambda}^{s}(\boldsymbol{k})=s \gamma / 2+\lambda \sqrt{\varsigma^{2}
k^{2}+(\Delta-s \gamma)^{2}/4}.  \label{eq1}
\end{equation}
Here,  $\boldsymbol{k}=\left(k_{x}, k_{y}\right)$ is the wave vector for an electron along MoS$_2$ sheet, $s=\pm1$ refers to the spin-up and spin-down respectively. The SOC parameter is $\gamma=75$ meV, $\Delta=1.66$ eV is the direct band gap energy in MoS$_2$ \cite{r14}. $\lambda=+ (-)$ represents the conduction (valence) bands respectively. $\varsigma=at$ with $a=3.193$ {\AA} being the lattice constant\cite{r15}. $t=1.1$ eV is the nearest neighbor hopping parameter. The wave function of electrons is given by \cite{r9}
\begin{equation}  
\psi_{\lambda, k}^{s}(r)=N_{\lambda}^{s}(\boldsymbol{k})\left[\varsigma k e^{-i \phi}
  / B_{\lambda}^{s}(\boldsymbol{k}), 1\right] e^{i k \cdot r},  \label{eq2}
\end{equation}
where $r=(x, y)$, $\quad B_{\lambda}^{s}(\boldsymbol{k})=E_{\lambda}^{s}(\boldsymbol{k})-\Delta
/ 2$, $\quad N_{\lambda}^{s}(\boldsymbol{k})=\left|B_{\lambda}^{s}(\boldsymbol{k})\right|
/\left[\left(B_{\lambda}^{s}(\boldsymbol{k})\right)^{2}+\varsigma^{2} k^{2}\right] $, and $\phi$ is the angle between $\boldsymbol{k}$ and the $x$-axis. Plasmons are generated by the collective excitations of electrons, which occur due to interactions between electrons through Coulomb forces.

The large band gap leads to significant influence from the Landau damping effect on interband collective excitations, originating from single-particle excitations \cite{r16}. Compared with the interband plasmon mode with relatively higher energy in the visible to the ultraviolet range, the energy of the intraband plasmon mode is lower, which is in the terahertz to the infrared range and is non-damped\cite{r6, r8, r9, re, kk}. This study focuses on intraband plasmons generated by electron transitions within the conduction band. These excitations involve collective oscillations of electrons, providing insights into the material's electronic properties under external perturbations. Under the random phase approximation (RPA) \cite{r17}, the dynamic dielectric function without considering phonon scattering is written as
\begin{equation}   
\varepsilon_{e e}\left(\boldsymbol{q}, \omega_{s}\right)=1-\left(V_{q} / 2\right)
\sum_{\boldsymbol{k}}\left(1+A_{\boldsymbol{k} \boldsymbol{q}}\right)
\Pi_{++}^{s}\left(\boldsymbol{k}, \boldsymbol{q}, \omega_{s}\right),  \label{eq3}
\end{equation}
where $A_{\boldsymbol{k}+\boldsymbol{q}}=(k+q \cos\theta) /|\boldsymbol{k}+\boldsymbol{q}|$, $\boldsymbol{q}$ is the momentum change during electron-electron scattering, $\theta$ is the angle
between $\boldsymbol{k}$ and $\boldsymbol{q}$, $V_q=2\pi e^2/(\varepsilon_\infty q)$is the 2D Fourier transform coefficient of the Coulomb interaction potential, $\varepsilon_\infty=5$ is the high-frequency dielectric constant of M-MoS$_2$ \cite{r6}.
\begin{equation}   
\Pi_{++}^{s}\left(\boldsymbol{k}, \boldsymbol{q}, \omega_{s}\right)=\frac{f\left[E_{+}^{s}(\boldsymbol{k}+\boldsymbol{q})\right]
-f\left[E_{+}^{s}(\boldsymbol{k})\right]}{\hbar
\omega_{s}+E_{+}^{s}(\boldsymbol{k}+\boldsymbol{q})-E_{+}^{s}(\boldsymbol{k})+i
\delta}  \label{eq4} 
\end{equation} 
is the density-density correlation function, with $f\left[E_{+}^{s}(\boldsymbol{k})\right]$ being the energy distribution function for electrons in different spin conduction bands. In the presence of optic-phonon scattering, the dielectric function can be obtained by using the self-consistent field theory
\cite{r18} and written as
\begin{equation} \begin{aligned}  
\varepsilon_{e p}(\boldsymbol{q}, 
\Omega_{s}) &=1-\sum_{\boldsymbol{k}}[\left(V_{q} / 2\right)
(1+A_{\boldsymbol{k},
\boldsymbol{q}}) \\&+ D_{0}(\omega_{q},
\Omega_{s})|U(\boldsymbol{k}, \boldsymbol{q})|^{2}]
\Pi_{++}^{s}\left(\boldsymbol{k},
\boldsymbol{q}, \Omega_{s}\right).   \label{eq5}
\end{aligned}
\end{equation}

Here $D_{0}\left(\omega_{q}, \Omega_{s}\right)=2 \hbar \omega_{q} /[\left(\hbar \Omega_{s}\right)^{2}-\left(\hbar \omega_{q}\right)^{2}] $ is the bare phonon propagator, $\Omega_s$ is the excitation frequency for different spin modes and $\omega_q$ is the optic-phonon frequency. Electrons in M-MoS$_2$
interact strongly with longitudinal optic-phonon (LO) via Fr\"{o}hlich coupling\cite{r19,r20}. In the long-wavelength approximation, the frequency of the polar LO phonon $\omega_q\simeq\omega_0$, and the energy of the polar LO phonon is $\hbar\omega_0 = 48$ meV, $ |U(\boldsymbol{k}, \boldsymbol{q})|^{2}=\left[g_{F} \operatorname{erfc}(q \sigma / 2)\right]^{2} $ is the effective electron-phonon interaction
matrix element, where $erfc(x)$ is the complementary error function \cite{r19,r21}, $g_F=98$ meV is the coupling constant, $\sigma=4.41$ {\AA} is the effective layer thickness.

When subjected to a driving electric field, electrons can be accelerated or even heated, causing the electron distribution to deviate from its original equilibrium state. As a result, commonly used equilibrium distribution functions, such as the Fermi-Dirac distribution and the Maxwell distribution, may no longer be applicable. To address this issue practically, a modified equilibrium distribution function can be employed to replace the non-equilibrium distribution function. This involves expressing the distribution function as a non-equilibrium energy distribution function \cite{r22}
$f\left[E_{+}^{s}(\boldsymbol{k})\right]\simeq f\left[E_{+}^{s}(\boldsymbol{k^*})\right]$,
$\boldsymbol{k^*}=\boldsymbol{k}+\boldsymbol{k_v}$ with $\boldsymbol{k_v}=m^*\boldsymbol{v}/\hbar$ being the momentum change induced by the electric field, $m^*=0.45 m_e$ is the effective mass \cite{r23}, $\boldsymbol{v}=(v_x,0)$ is the drift velocity of electrons generated by the electric field. $f(E)=[e^{(E-\mu) / k_{B} T_{e}}+1]^{-1} $ is the modified Fermi-Dirac distribution function, where $T_e$  is the temperature of the electron system under the electric field. Considering that at 0 K or extremely low temperatures, the chemical potential of electrons $\mu\simeq E_F$, we directly substituted the value of the Fermi wave vector $k_F= \sqrt{4\pi n_e}$ into Eq. \eqref{eq1} to obtain the Fermi energy, i.e. the chemical potential $\mu=s \gamma/2+\lambda \sqrt{\varsigma^{2} k_F^{2}+(\Delta-s \gamma)^{2}/4}$, $n_e$ is the density of electrons in the conduction band. Here, the parameters used $v_x$ and $T_e$ can be found in references \cite{r21,r24}, which can be obtained by self-consistent calculations.

In this work, the application of the driving electric field not only affects the eigenenergy of electrons in ML-MoS$_2$ but also impacts the electron wave functions. On one hand, the polarization effect of the electric field modifies the distribution of electrons, and the screening effect of electrons is affected. This leads to the change in the overlap of electron wave functions, thereby influencing the many-body interaction behaviors of electrons, including the collective excitations of electrons i.e. plasmons.
Regarding this aspect, we haven't found relevant research in the literature, and it can be the next work focus. On the other hand, under the action of the driving electric field, the energy band structure of electrons will tilt, and the electron energy will change. From a statistical perspective, the distribution of electrons is no longer in an equilibrium state. Therefore, we use a non-equilibrium energy distribution function $f\left[E_{+}^{s}(\boldsymbol{k^*})\right]$ to describe the distribution of electrons under an electric field\cite{r22}. Therefore, here the influence of the electric field on plasmons originating from the collective excitations of electrons is mainly considered from a statistical perspective.

The plasmon mode without phonon scattering can be obtained by Re$[\varepsilon_{ee}\left(\boldsymbol{q},
\omega_{s}\right)] \rightarrow 0$. In the long-wavelength approximation ($\lim q\rightarrow 0$), the frequency of plasmon under the driving electric field can be obtained as
\begin{equation}  \begin{aligned}  
\left(\hbar \omega_{s}\right)^{2}&=\frac{4 e^{2} q}
{\varepsilon_{\infty} \pi} \int_{0}^{k_{F}} \int_{-\pi}^{\pi}d\phi dk~[f(k)k]
\\&\times \frac{\left(\varsigma^{4} k^{2} \sin (\varphi-\phi)^{2}+\varsigma^{2}
 (\Delta-s \gamma)^{2} / 4\right)}{\left(\varsigma^{2} k^{2}+(\Delta-s \gamma)^{2}
 / 4\right)^{3 / 2}}, \label{eq6}
 \end{aligned} 
 \end{equation}
where $\varphi$ is the angle between the plasmon wave vector $\boldsymbol{q}$ and $x$ axis. In the presence of phonon scattering, the coupled plasmon-phonon mode is given by Re$[\varepsilon_{ep}\left(\boldsymbol{q}, \Omega_{s}\right)]\rightarrow0$, and in the long-wavelength approximation, the coupled plasmon-phonon mode under the driving electric field can be written as
\begin{equation} \begin{aligned}  
\Omega_{s}&=\frac{1}{\sqrt{2}}\Big[\omega_{q}^{2}+\omega_{s}^{2}
\\&\pm \sqrt{\left(\omega_{q}^{2}-\omega_{s}^{2}\right)^{2}+4 g_{F}^{2} \omega_{q} \omega_{s}^{2}
  \varepsilon_{\infty} \sigma /\left(e^{2} \hbar \pi\right)}~~\Big]^{1/2},  \label{eq7}
\end{aligned}
\end{equation}
where $\omega_s$ is the frequency of plasmons under the driving electric field given by Eq. \eqref{eq6}.
In this study, we do not consider the weak electric field approximation. Therefore, the result is universal for TMDs. 

\section{Results and Discussions}
\begin{figure}
\setlength{\abovecaptionskip}{0.1cm}
\includegraphics[width=0.40\textwidth,height=0.20\textheight,angle= 0]{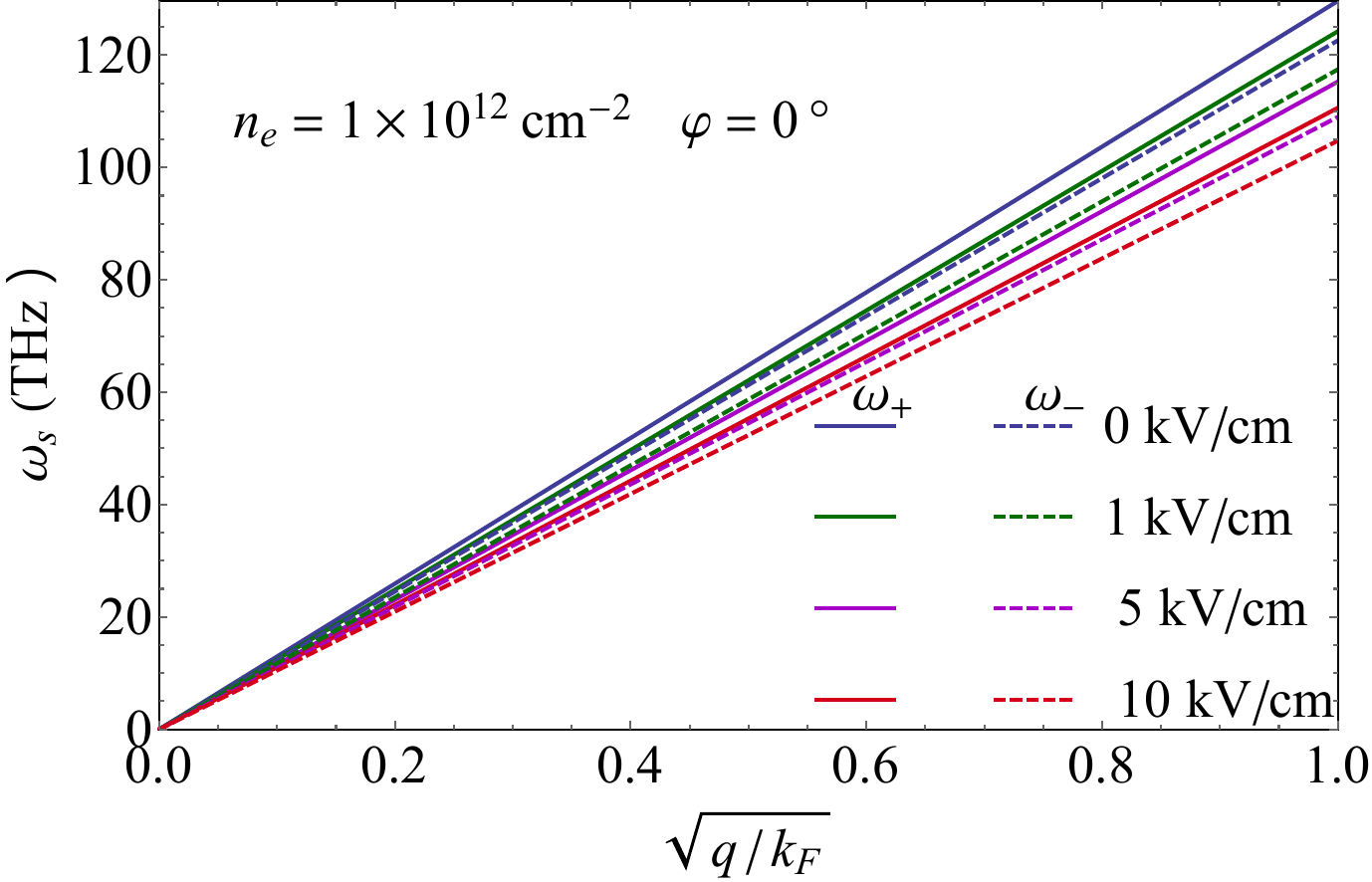}
\caption{The dispersion relation of spin up (solid line) and spin
down (dashed line) plasmon modes for different electric fields at
the fixed electron density and angle $\varphi$.} \label{fig1}
\end{figure}

There are two tunable plasmons in the terahertz region, resulting from SOC in conjunction with electron-electron interactions, as described in Eq. \eqref{eq6}. Figure \ref{fig1} illustrates the dispersion of plasmons from two spin conduction bands under varying electric fields. The plasmon mode is in the terahertz to infrared range, which is the same as the situation without an electric field applied \cite{r6, r8, r9, re, kk}. The electric field causes a drift in the electron distribution, leading to a change in plasmon frequency. For all wave vectors, the frequencies of both spin-up and spin-down plasmon modes decrease with an increasing electric field. Although spin-orbit coupling is usually considered to affect the band gap and the interband behavior of electrons, such as excitons exciting. The many-body interaction process of electrons is related to their energy and momentum. Therefore, spin-orbit coupling will have an impact on both the inter-band and intra-band behavior of electrons. Compared with the inter-band behavior, the intra-band behavior is relatively less affected. In this paper, we study the collective excitation resulting from electron-electron interactions within different spin conduction bands. As can be seen from Figure 1, spin-orbit coupling indeed leads to differences in the plasmons of two different spin modes. The frequency of spin-up plasmon modes is consistently higher than that of spin-down plasmon modes for all examined electric fields. Similar to the behavior of plasmons in graphene subjected to driving electric fields \cite{r25}, the frequency of plasmons in TMDs continues to depend on $q^{1/2}$ when an electric field is applied. The results depicted in the figure indicate that the frequency of plasmons in ML-MoS$_2$ can be controlled by applying a driving electric field. In this study, we used the most commonly used electron density $10^{12}$ cm$^{-2}$ when studying the transport properties of ML-MoS$_2$\cite{r6,br}. From the previous research work\cite{r9}, it can be seen that as the electron density increases, the frequency of the plasmon increases. From our current results, it can be observed that as the frequency of the plasmon increases, the modulation of the electric field is significantly enhanced. Therefore, as the electron density increases, the influence of the electric field on the plasmon becomes greater. Conversely, the smaller the electron density, the smaller the influence. However, it should be noted that our work has studied the plasmon modes by using the dielectric function theory under the RPA. We know that the RPA is suitable for the situation where the electron density is high and the interaction between electrons is weak. Therefore, it is physically meaningless to discuss the cases with a lower electron density of less than $10^{11}$ cm$^{-2}$.

\begin{figure}
\setlength{\abovecaptionskip}{0.1cm}
\includegraphics[width=0.38\textwidth,height=0.20\textheight,angle= 0]{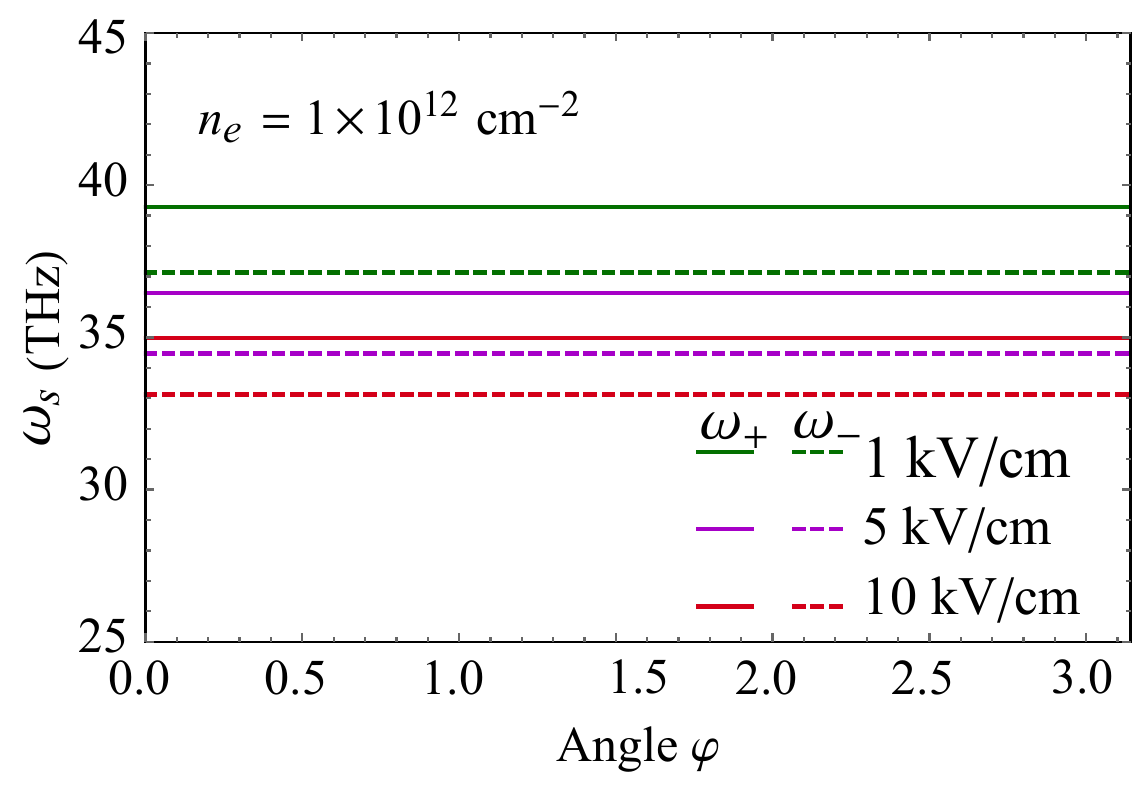}
\caption{The angular dependence of frequencies of the spin up (solid line) and spin
down (dashed line) plasmon modes at a fixed electron density and $q=0.1k_F$ for different driving electric fields. }
\label{fig2}
\end{figure}

The applied driving electric field breaks the symmetry of the system. As shown in Eq. \eqref{eq6}, the frequency of the plasmons depends on the angle $\varphi$, which is the angle between the wave vector $\boldsymbol{q}$ and the x-axis. Figure \ref{fig2} illustrates that the frequency of the plasmon decreases as the electric field increases, regardless of the angle. Furthermore, both plasmon modes exhibit a very weak dependence on the angle $\varphi$ under the electric fields, making any variation in plasmon frequency almost imperceptible. The small coefficient in front of $\sin(\phi-\varphi)$ in Eq. \eqref{eq6} contributes to the minimal dependence of frequency on the angle $\varphi$. This indicates that the driving electric field does not introduce significant anisotropy to the plasmon modes in TMDs. Understanding this behavior is essential for comprehending the optoelectronic properties of 2D TMDs in electric fields, particularly at high field strengths. From a physical standpoint, the relationship between the frequency of the plasmon and the angle is governed by the conservation of energy and momentum during electron-electron interactions.

In this work, the influence of the dielectric wafer on the plasmons in M-MoS$_2$ was not taken into account. In fact, the presence of the dielectric wafer can affect the collective excitation behavior. On the one hand, the high-frequency dielectric function of M-MoS$_2$ can be replaced by an effective dielectric function to reflect this influence due to the mismatch of the dielectric constants at the interface of the structure\cite{eh}. Such as for M-MoS$_2$ on a SiO$_2$ substrate, the effective dielectric function should be $(\varepsilon_\infty+\varepsilon_{SiO_2})/2=4.5$, where $\varepsilon_{SiO_2}=4$ is the dielectric constant of SiO$_2$. This is slightly smaller than the high-frequency dielectric function of $\varepsilon_\infty=5$ used for M-MoS$_2$ in this work. Therefore, if the dielectric effect of the dielectric wafer is considered, the plasmon frequency in M-MoS$_2$ should be slightly higher than the current value. However, this does not change the behavior of the plasmon mode under the electric field. On the other hand, at relatively low temperatures, the charged impurities in the dielectric wafer can affect the plasmon frequency. Impurity scattering can enhance the screening effect of electrons, leading to a decrease in the plasmon frequency. This is a typical result in one and two-dimensional electronic systems, where impurity scattering can strongly affect the dispersion and damping of plasmons\cite{sd,ig}. In addition, the optical phonons of the dielectric wafer can also couple with plasmons to form a coupled mode\cite{r9}. As for the weak influence of the electric field on the anisotropy of plasmons in this article, it stems from the properties of M-MoS$_2$ itself, that is, the small coefficient in front of $\sin(\phi-\varphi)$ in Eq. \eqref{eq6} leads to this. If an anisotropic dielectric wafer is introduced and the coupling between the dielectric wafer and M-MoS$_2$ is considered, its anisotropic dielectric effects, such as an anisotropic dielectric function, will directly lead to the anisotropy of the plasmon modes, but this is not caused by the applied electric field, and the anisotropy caused by the electric field will not be significantly changed.

\begin{figure}
\setlength{\abovecaptionskip}{0.1cm}
\includegraphics[width=0.40\textwidth,height=0.20\textheight,angle= 0]{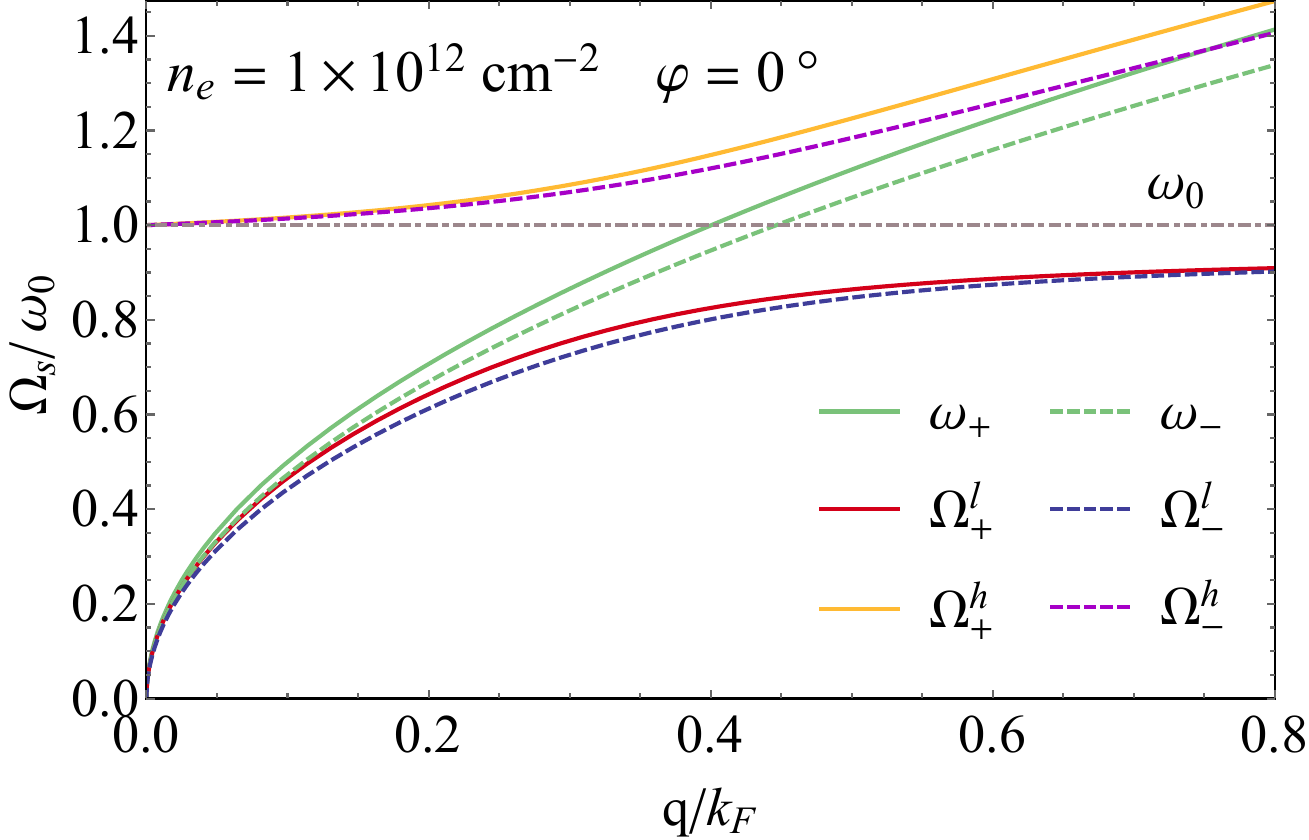}
\caption{The plasmon-phonon coupling modes at a fixed electron density for the angle $\varphi=0^\circ$ and the electric field strength $F_x=10$ kV/cm. }
\label{fig3}
\end{figure}

Figure \ref{fig3} depicts the coupled plasmon-phonon modes $\Omega_{+}^{h}$, $\Omega_{+}^{l}$, $\Omega_{-}^{h}$, and $\Omega_{-}^{l}$ respectively. These modes originate from the coupling of the optic-phonon mode $\omega_0$ with the plasmon modes $\omega_{+}$ and $\omega_{-}$, which are derived from the spin-up and spin-down branches respectively. It is observed that plasmons can strongly couple with the optic phonon when their energies are in the vicinity of $\omega_0$, representing the energy crossover between plasmons and optic phonons. At the resonant point, Rabi-splitting and anticrossing phenomena occur, giving rise to the emergence of two distinct coupled plasmon-phonon modes. Conversely, when the energies of the coupled modes deviate significantly from the resonance point, these modes decouple, reverting to either the plasmon mode or the phonon mode. As presented in Figure \ref{fig3}, for the high-frequency coupled branches $\Omega_{+}^{h}$ and $\Omega_{-}^{h}$, they exhibit phonon-like characteristics for small wave vector values $q$, but transform into plasmon-like behavior as $q$ increases. In contrast, the low-frequency coupled branches $\Omega_{+}^{l}$ and $\Omega_{-}^{l}$ display the opposite trend. Furthermore, the coupled high-frequency branches $\Omega_{+}^{h}$ and $\Omega_{-}^{h}$ demonstrate a more pronounced dependence on the electric field. Significantly, this indicates that the external electric field can control the occurrence of Rabi-splitting and anticrossing. Given that plasmons and phonons are the most fundamental excitations, these results hold great potential for the development of 2D TMDs-based optoelectronic devices. As shown in Figure \ref{fig1}, the frequency of plasmons decreases with an increase in the electric field. Consequently, the wave vector of the plasmon, whose energy matches that of the optic phonon, increases with the electric field. Moreover, the crossing point between the plasmon and optic-phonon energies shifts towards larger wave vectors as the electric field strength increases. These findings suggest that the driving electric field can exert a substantial influence on the interaction between plasmons and optic-phonon.

\begin{figure}
\includegraphics[width=0.48\textwidth,height=0.18\textheight,angle= 0]{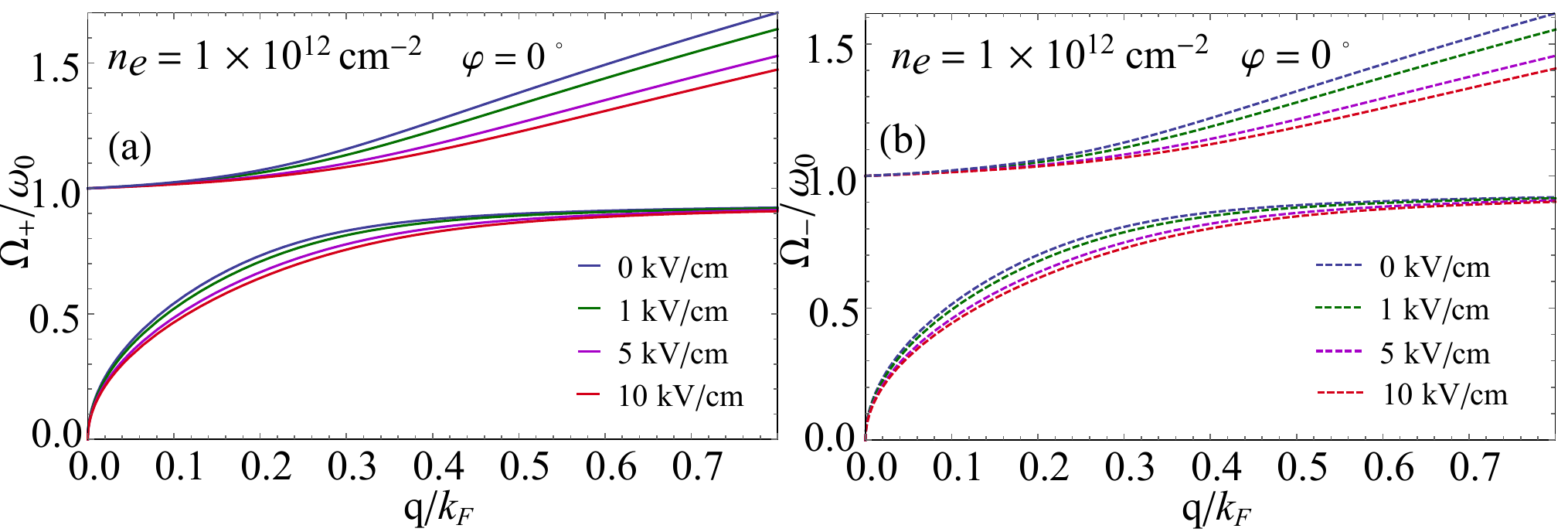}
\caption{The dispersion relation for coupled spin-up (a) and
spin-down (b) plasmon-phonon modes for different electric fields at
a fixed electron density and $\varphi=0^\circ$. } \label{fig4}
\end{figure}

Figure \ref{fig4} meticulously portrays the behavior of coupled plasmon-phonon modes under the driving electric fields. As the intensity of the driving electric field is incrementally augmented, the frequencies of both the high-frequency and the low-frequency coupled branches undergo a consistent decline. This phenomenon can be ascribed to the concomitant reduction in the frequencies of uncoupled plasmon modes as the electric field strength escalates, as vividly demonstrated in Figure \ref{fig1}. The uncoupled plasmon, which serves as a fundamental building block for the formation of coupled plasmon-phonon, is directly affected by the applied electric field. When the electric field interacts with the electron gas responsible for plasmon generation, it modifies the electron-electron and electron-phonon interactions. This, in turn, leads to a decrease in the characteristic frequencies of the uncoupled plasmon modes. Since the coupled plasmon-phonon results from the hybridization of plasmon and phonon modes, the change in the frequencies of the plasmons is propagated to the coupled modes, thereby causing the observed reduction in the frequencies of both the high- and low-frequency coupled branches. Moreover, the angular dependence of the plasmon-phonon coupling is a crucial aspect that merits in-depth examination. Given that the coupled plasmon-phonon modes are formed through the interaction of plasmon and phonon modes, and the plasmon component has a weak angular dependence, it can be logically inferred that the coupled plasmon-phonon modes also display a similar weak dependence on $\varphi$. Specifically, across all possible angles $\varphi$ within the 2D system, the frequency of the coupled plasmon-phonon modes consistently decreases with an increasing electric field. This behavior is in line with the trend observed for uncoupled plasmon modes, underscoring the fact that the driving electric field exerts a dominant and consistent influence on the collective excitation modes, regardless of the angular orientation. These findings not only deepen our fundamental understanding of the physical mechanisms governing the interaction between plasmons and phonons in the presence of an external electric field but also have far-reaching implications for the design and optimization of advanced optoelectronic devices based on 2D materials. In such devices, precise control over these collective excitation modes is essential for achieving desired optical and electrical properties, and the insights gained from this study can guide the development of novel materials and device architectures. Although no relevant experimental research has been found on this topic currently. We hope that our research can lay the theoretical foundation for experiment works and be verified by them. We propose that polarization-dependent spectroscopic measurements and scanning near-field optical microscopy (SNOM) offering nanoscale spatial resolution can be used to discover our theoretical findings.

In this work, we did not consider the damping of these intra-band plasmons. This is because, for ML-MoS$_2$, the intra-band plasmon modes with lower energy have no damping, while the inter-band plasmon modes with higher energy have strong damping due to their coupling with excitons \cite{r16}. The damping of plasmons originates from the decay of the collective excitation mode of plasmons into individual excitations. The damping region can be obtained from the imaginary part of the dielectric function \cite{r6}. Considering that the imaginary part of the dielectric function is also related to the electron distribution, the drift of electrons and the nonequilibrium distribution of electrons under the driving electric field will also affect the damping of inter-band plasmons. Although there is no research on this in the works of literature yet, we can conduct a qualitative analysis. The driving electric field causes the energy bands of electrons to tilt, changes the electron energy distribution, makes the electron distribution drift towards the lower energy region, lowers the Fermi surface, weakens the scattering interaction between electrons, and thus reduces the damping effect of inter-band plasmons.

\section{Conclusion}
In conclusion, this study examines the effect of electric fields on plasmons and coupled plasmon-phonons in monolayer TMDs, considering the SOC. The key findings are: (i) The frequencies of plasmons and coupled plasmon-phonons from the two spin-split conduction bands decrease as the electric field increases, regardless of the wave vector. (ii) The frequencies of the plasmons and coupled plasmon-phonons are not sensitive to the angle of the plasmon under a strong electric field. (iii) An applied electric field can effectively control the coupling between plasmons and optical phonons. (iv) The frequencies of the high-frequency branches are more strongly affected by electric fields. These results indicate that plasmons and coupled plasmon-phonons can be tuned not only by modifying the gate voltage to regulate electron concentration but also by applying an electric field. The findings can be employed to investigate the experimental performances of optoelectronic devices based on TMDs, particularly when subjected to intense electric fields.

\begin{acknowledgments} This research was funded by the Fundamental Research Program of Shanxi Province, China (No. 202103021224250), the Science and Technology Innovation Project of Colleges and Universities of Shanxi Province of China (No. 2020L0242), the Higher Education Reform and Innovation Project of Shanxi Province (No. J20230617).
\end{acknowledgments}

\section*{AUTHOR DECLARATIONS}
{\bf Conflict of Interest:} The authors have no conflicts to disclose.

\section*{DATA AVAILABILITY}
The data that support the findings of this study are available from the corresponding authors upon reasonable request.


\end{document}